\documentclass[10pt, conference]{IEEEtran}
\usepackage{nopageno}
\IEEEoverridecommandlockouts
\UseRawInputEncoding

\usepackage{cite}
\usepackage{amsmath,amssymb,amsfonts}
\usepackage{algorithmic}
\usepackage{graphicx,subcaption,epstopdf}
\usepackage{textcomp}
\usepackage{xcolor}
\usepackage{tabularx,ctable}
\def\BibTeX{{\rm B\kern-.05em{\sc i\kern-.025em b}\kern-.08em
    T\kern-.1667em\lower.7ex\hbox{E}\kern-.125emX}}
    
\usepackage{fancyhdr}

\begin{document}

\title{Medium Access Strategies for Integrated Access and Backhaul at mmWaves Unlicensed Spectrum}

\author{\IEEEauthorblockN{Biswa P. S. Sahoo\IEEEauthorrefmark{1}, Satyabrata Swain\IEEEauthorrefmark{2}, Hung-Yu Wei\IEEEauthorrefmark{1}, and Mahasweta Sarkar\IEEEauthorrefmark{3}
‬ }
\IEEEauthorblockA{
\IEEEauthorrefmark{1}Department of Electrical Engineering, National Taiwan University, Taipei, Taiwan \\
\IEEEauthorrefmark{2}Department of Information Technology, Vellore Institute of Technology, Vellore, Tamil Nadu, India \\
\IEEEauthorrefmark{3}Department of Electrical and Computer Engineering, San Diego State University, San Diego, California, USA}
}

\maketitle

\begin{abstract}
The unlicensed spectrum is recently considered one of the deﬁning solutions to meet the steadily growing trafﬁc demand. This, in turn, has led to the enhancement for LTE in Release-13 to enable Licensed-Assisted Access (LAA) operations. The design of the medium access control (MAC) protocol for the LAA system to harmonically coexist with the incumbent WLAN system operating in an unlicensed band is critical and challenging. In this paper, we consider an Integrated Access and Backhaul (IAB) system coexisting with a Wi-Fi network operating at millimeter-wave (mmWave) unlicensed spectrum, for which a listen-before-talk-based (LBT) based medium access mechanism is carefully designed. Additionally, we have considered an in-band system that supports both access and backhaul in a single node where the small-cell or the IAB nodes compete with the WiGig for medium access. We present comprehensive experimental results and give design insights based on the simulation results.
\end{abstract}


\IEEEpeerreviewmaketitle
\thispagestyle{fancy}
\pagenumbering{gobble}

\section{Introduction}
In recent years, the number of wireless devices and broadband subscriptions has continued to grow steadily. This, in turn, has led to a signiﬁcant increase in the capacity requirements, which led to the evolution of ﬁfth generation (5G) technologies~\cite{pi2011introduction, sahoo2017millimeter, sahoo2018enabling}. There are broadly two ways of squeezing more capacity out of mobile radios: more spectrum or better spectral efﬁciency (more bits per Hz). The recent Federal Communications Commission (FCC) regulation update releases about 14 GHz spectrum for unlicensed use in the 57-64 GHz band~\cite{chen2020impact}. This creates a massive platform for cellular operators for trafﬁc ofﬂoading from the cellular network onto this band. Thus, the use of unlicensed spectrum in long-term evolution (LTE) has received much attention recently and also has been endeavoring by the 3rd Generation Partnership Project (3GPP) as the featured candidate technology to deliver cellular services~\cite{cui2020learning}. For making the best use of the unlicensed band, multiple technologies will coexist for different needs such as enterprises, small businesses, venues, and residential/neighborhood.

The coexistence of LAA and Wi-Fi in the unlicensed band has drawn extensive attention in both industry and academia. Several works have recently investigated the coexistence mechanism for LTE and Wi-Fi over sub-6 unlicensed bands (e.g., 5GHz 0r 2.4 GHz). The 3GPP has investigated the feasibility of Integrated Access and Backhaul (IAB) system as part of its Rel-15 standardization activities~\cite{yao2017outage,lei2020deep,inoue20205g,zhai2020mesh,saha2019load,polese2020integrated}. The objective of IAB is to design an advanced multi-hop wireless backhaul relay that ensures efficient self-backhauling in New Radio (NR) base stations\cite{saha2019millimeter}. This paper investigates the scheduling policies of an in-band self-backhauled unlicensed network, referred to as NR-U, throughout this paper. In an NR-U system, before attempting any transmission, by regulatory requirement, the node needs to perform a clear channel assessment (CCA) to check if the channel is idle or busy, which is also called LBT. The node transmits only if the medium is sensed idle. Generally speaking, for an IAB node, a straight-forward solution could be to time-multiplexed the backhaul link and access link and allocate time domain resources in their respective slot. However, this solution sometimes may end up wasting the allocated timeslots due to unsuccessful medium access. Thus, it is imperative to consider the LBT and in-band resource allocation jointly.
  
NR-U has been investigated in a 3GPP Rel-16 WI, which enables its inclusion in future NR specifications. Extending the NR-based access across different Radio Access Technologies (RATs) adhering to LBT requirements for a fair coexistence in unlicensed spectrum is crucial. When NR-U is applied to an IAB system for in-band self-backhauling, allocating time resources, it is required to adhere to meet LBT requirements. However, jointly addressing the LBT as well as in-band resource allocation is challenging. Therefore, any fixed pattern resource allocation strategy for in-band scheduling will not serve the purpose. Taking this into consideration, in this paper, we proposed a flexible scheduling strategy to utilize the radio resources across different multiple RATs effectively. As a downside of LBT, it may increase the access delay to affect the total latency. We will handle this by introducing a centralized controller to the LBT strategy.

The rest of this paper is organized as follows. Section~\ref{sec:sys_over} presents the system overview and describes the problem. In Section~\ref{sec:pros_soln}, we provide three medium access strategy in detail. Section~\ref{sec:per_eval} numerical results are presented and analyzed. Finally, conclusions are drawn in Section~\ref{sec:conclusion}.

\section{System Overview}\label{sec:sys_over}
In this section, we provide the details of the investigated system model, network traffic, and mmWave communication channel with multi-antenna transceivers suitable for 5G NR systems.

\subsection{Network, Connections, and Traffic}
\begin{figure}[t]
    \centering
    \includegraphics[width=\columnwidth]{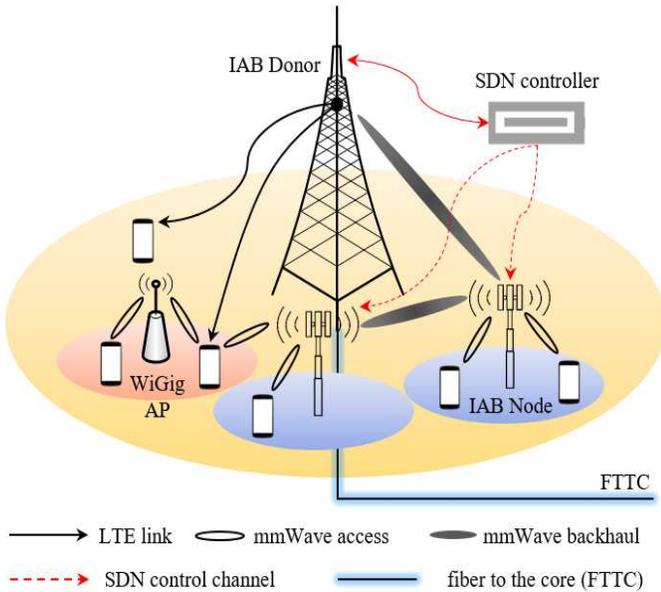}
    \caption{Network consisting of one IAB donor, two IAB nodes, one Wi-Gig AP, and multiple UEs which could conveniently connect to either of the RATs. For the purpose of simplicity, the beam management is taken care by SDN controller.}
    \label{fig:network}
\end{figure}

We consider a typical two-tier IAB network consisting of an IAB donor, a set of IAB nodes, a set of WiGig access points (APs), and a set of user equipment (UEs) that could conveniently connect to either of the three RATs at a time.  Let $\mathcal{B} = \{b_0, b_1, b_2, \dots b_B\}$ denote the set of all IAB nodes including the IAB donors. The set of UE associated with IAB nodes is denoted as $\mathcal{U} = \{u_0, u_1 \ldots, u_U\}$. The IAB donor is connected to the core network by high-speed fiber to the cell (FTTC) links, provides a root from which connections go to the IAB node at the edge via relay IAB node that is self-backhauled on mmWave band.  We assume that IAB nodes are deployed inside a circular cell of radius $R$ with the IAB donor at its center. The IAB-donor could operate in both licensed and mmWave unlicensed bands, whereas IAB nodes and WiGig APs share the same mmWave unlicensed band by transmitting in different fractions of time via some medium access strategy. However, for the simplicity of implementation, we have ignored the IAB-donor operation mechanism. Fig.~\ref{fig:network} illustrates the network model.

We refer to the link between a UE and IAB node or AP as an \emph{access link} and the link between an IAB donor and IAB node as a \emph{backhaul link}. We assume that the same mmWave band is shared by both access and backhaul links. Admissible connections are IAB donor $\rightleftharpoons$ IAB node, IAB donor $\rightleftharpoons$ UE, IAB node $\rightleftharpoons$ IAB node, IAB node $\rightleftharpoons$ UE, AP $\rightleftharpoons$ UE, with both downlink and uplink traffic flows. We assume that packet arrival to the WiGig AP and the IAB-nodes follows a Poisson process with the average arrival rate as $\lambda$ packets per sec.

\subsection{Medium Access for Wi-Fi and IAB}
The coexistence of different RATs primarily relies on the physical propagation characteristics and channel access mechanism. For example, mmWave communication imposes beamforming and directional transmissions to overcome propagation limits, which play a significant role during medium access procedure. While the commercial network operators try to access the unlicensed spectrum, it is essential to ensure a fair and harmonious coexistence with the incumbent unlicensed systems. Here, we assume the IAB donor has sufficient resources and can provide reliable supports to UEs when the unlicensed band is unavailable.

\begin{figure}[t]
    \centering
    \includegraphics[width=\columnwidth]{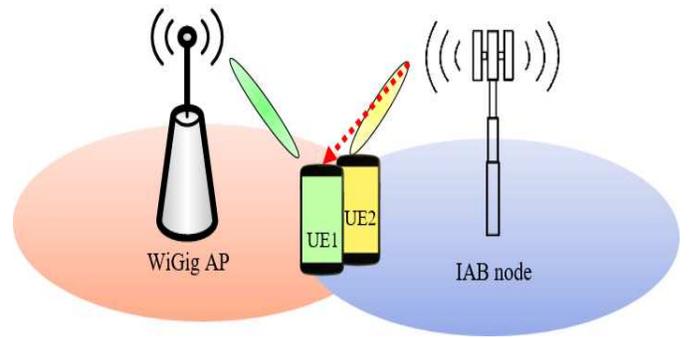}
    \caption{Interference problems due to incorrect LBT. UE1 is interfered by signal from IAB-node intended towards UE2.}
    \label{fig:inf}
\end{figure}

The WiGig devices (IEEE 802.11ad/ay) employ hybrid medium access, which compromises both contention-based access (CSMA/CA) and time-based access (TDMA)~\cite{standard201211}. However, the current firmware only supports CSMA/CA channel access. Thus, WiGig devices still encounter deafness issues, which did not exist in Omni-directional communications~\cite{wang2020unlicensed}. WiGig is the incumbent primary system in the 60 GHz unlicensed spectrum band. 3GPP recently adopted the listen-before-talk (LBT) mechanism to control the channel accesses for LTE/Wi-Fi coexistence. However, when UEs are closely located, as illustrated in Fig.~\ref{fig:inf}, the directional LBT might result in incorrect channel status and thus led to interference. The interference dynamics are very different at the transmitter and receiver sides.

We carefully designed the LBT mechanism by employing a centralized SDN controller to handle the receiver's interference due to incorrect LBT. 

\subsection{Software-Defined Networking (SDN) Controller}
We assumed an SDN controller that is implemented to manage the mmWave-enabled IAB mesh network. The SDN controller is connected to the IAB donor and IAB nodes, as shown in Fig.~\ref{fig:network}. The controller features basic network management capabilities, such as:
\begin{itemize}
\item \textit{IAB Configuration}: The SDN controller ensures the interference at the receiver side is minimized by re-configuring the mmWave transmitting link/beam or both. The channel status of UEs is constantly fed to the controller via the serving IAB node. Thus, the controller has a partial view of the UE channel status at the receiver side. We assume a mechanism is in place to feedback the channel status as in~\cite{sahoo2018sdn}. 

\item \textit{Beam Steering and Alignment}: The SDN controller also manages the directional beam alignment between the IAB node and between the IAB node and UE. It also manages the angle of arrival (AoA) and angle of departure (AoD).  
\end{itemize}

We introduced the SDN controller is to provide efficient resource coordination among serving IAB-nodes during user mobility and guard against the potential radio link failure while ensuring at least one of the RATs is connected to the UE at all times.

\subsection{Millimeter-Wave Channel Model}\label{sec:ch_mod}
The mmWave channel has limited scattering~\cite{hur2016proposal, ko2017millimeter, akdeniz2014millimeter}. Thus, we assume that the channel has $L$ taps, where $\ell = 0, \dots, L$ is the subpath index. Under this adopted channel model, the channel matrix $\mathbf{H}$ can mathematically be expressed as

\begin{equation}
\mathbf{H} = \sum_{\ell = 0}^{L} g_{\ell} \mathbf{u}_{rx}(\theta_{\ell}) \mathbf{u}_{tx}^{\text{H}}(\phi_{\ell})
\end{equation}
where $g_{\ell}$ is the complex small-scale fading coefficient associated with the $\ell$-th propagation path,  $\mathbf{u}_{rx}(\theta_{\ell})$ and $\mathbf{u}_{tx}(\phi_{\ell})$ are the antenna array response vectors with AoD $\phi_{\ell}$ and AoA $\theta_{\ell}$ of the $\ell$-th path at the transmitter and receiver, respectively. Assuming uniform linear arrays (ULAs), the antenna array response $\mathbf{u}_{rx}(\theta_{\ell})$ and $\mathbf{u}_{tx}^{\text{H}}(\phi_{\ell})$ are adopted as defined in \cite{wu2019efficient}.

Without loss of generality, we assume the mmWave communications based on the abstraction used in the prior studies, the received power $P_r$ at the receiver can be calculated as:
\begin{equation}
P_r (u, v) = P_t (u, v) \cdot \mu(u, v) \cdot \gamma^{-1} \cdot PL^{-1} 
\end{equation}
where $P_t(u, v)$ is a reference power or transmitted power, $\mu$ is the combined antenna gain of transmitter and receiver, $\gamma$ is the subpath attenuation, and $PL$ denotes the associated line-of- sight (LOS) pathloss in dB and can be derived as:
\begin{equation}\label{eq:pth_loss}
PL(d) (dB) = \alpha + 10\beta \cdot 10\log_{10}(d) + \eta
\end{equation}
where $PL(d)$ is the mean pathloss, over a reference Tx-Rx separation distance $d$, in dB, $\alpha$ is the floating intercept in dB, $\beta$ is the pathloss exponent, $\eta \sim N(0, \sigma^2)$. The simulated values are provided in Table~\ref{tbl:sys_par}.

We computed the backhaul throughput considering adjacent channel interference (ACI) as defined in~\cite{3GPP_ch_mod}. The system throughput, as expressed in (\ref{eq:outageThput}), is evaluated by Shannon capacity, which is dependent on the received SINR. Specifically, each user's serving beams will cause interference received by other users in the same transmission group.
\begin{equation}
Thput [bpshz] = \varphi \cdot \omega + \log_2\bigg(1 + \frac{P_r(u, v)}{N + I_{ACI}}\bigg)
\label{eq:outageThput}
\end{equation}
where $\varphi$ is the bandwidth overhead, $\omega$ is the mmWave system bandwidth, and $I_{ACI}$ is the adjacent channel interference (ACI), and $N$ is Additive white Gaussian noise (AWGN).


\subsection{Medium Access Probability}
The probability that a WiGig AP transmits a packet in a slot time which is borrowed from \cite{song2015coexistence} by solving the Markov model is obtained as follows
\begin{equation}\label{eq:wifi_access_prob}
    \mathbb{P}_\omega = \frac{2(1 - 2P_{cw})}{(1 - 2P_{cw})(C + 1) + P_{cw}C(1 - (2P_{cw})^m)}
\end{equation}
where $P_{cw}$ is the collision probability of Wi-Fi nodes, $C \sim [0, C - 1]$ is maximum backoff counter, and $m$ is the total number of Wi-Fi APs. Similarly, the probability that a IAB node accessing the channel to transmit in a time slot is calculated from \cite{song2015coexistence} by solving the Markov model is obtained as follows
\begin{equation}\label{eq:iab_access_prob}
    \mathbb{P}_\Delta = \frac{\frac{1}{Z}P_{cb}\sum_{j=1}^{Z}(1-P_{cb})^{j-1}}{1 - \frac{1}{Z}(1 - P_{cb})\sum_{j=1}^{Z}(1-P_{cb})^{j-1}}
\end{equation}
where $P_{cb}$ is the constant and independent collision probability of IAB nodes, $Z \sim [0, Z-1]$ is the CW size of the IAB nodes.

\section{Proposed Medium Access Strategy}\label{sec:pros_soln}
\begin{figure*}[t]
    \centering
    \begin{subfigure}[t]{0.3\textwidth}
        \centering
        \includegraphics[height=1.2in]{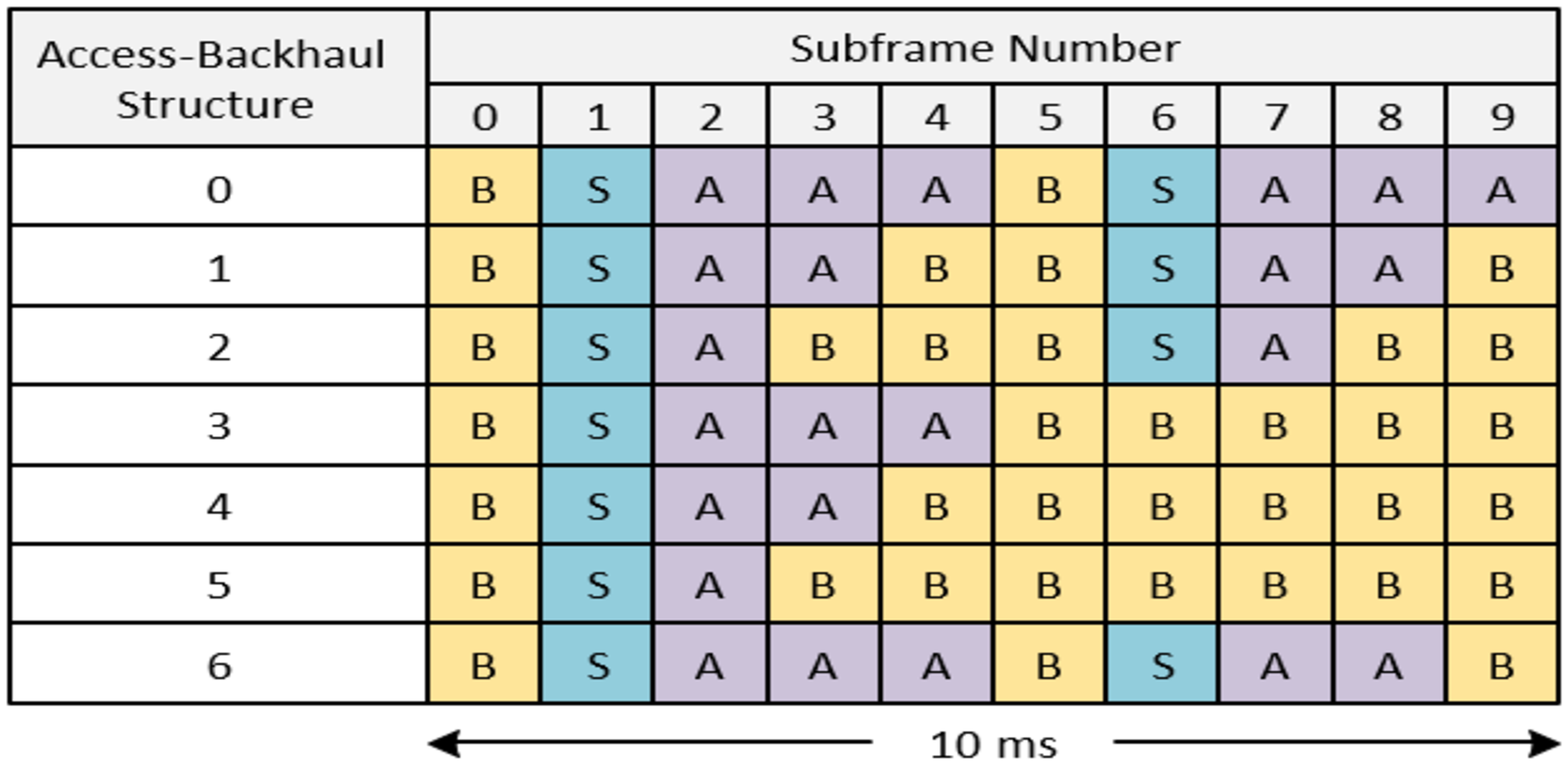}
        \caption{Baseline Strategy}
    \end{subfigure}%
    ~ 
    \begin{subfigure}[t]{0.3\textwidth}
        \centering
        \includegraphics[height=1.2in]{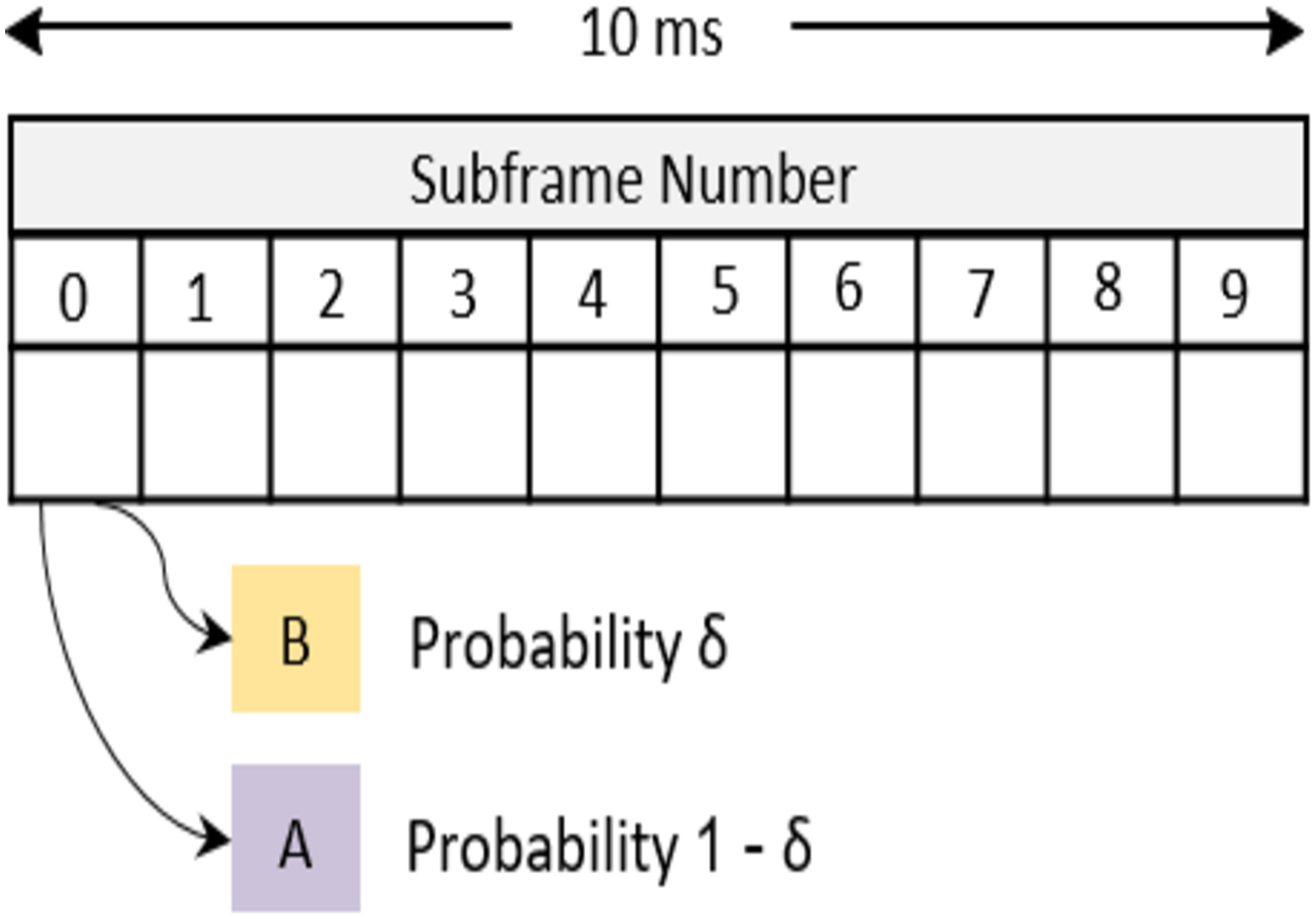}
        \caption{Probabilistic Strategy}
    \end{subfigure}%
    ~ 
    \begin{subfigure}[t]{0.3\textwidth}
        \centering
        \includegraphics[height=1.2in]{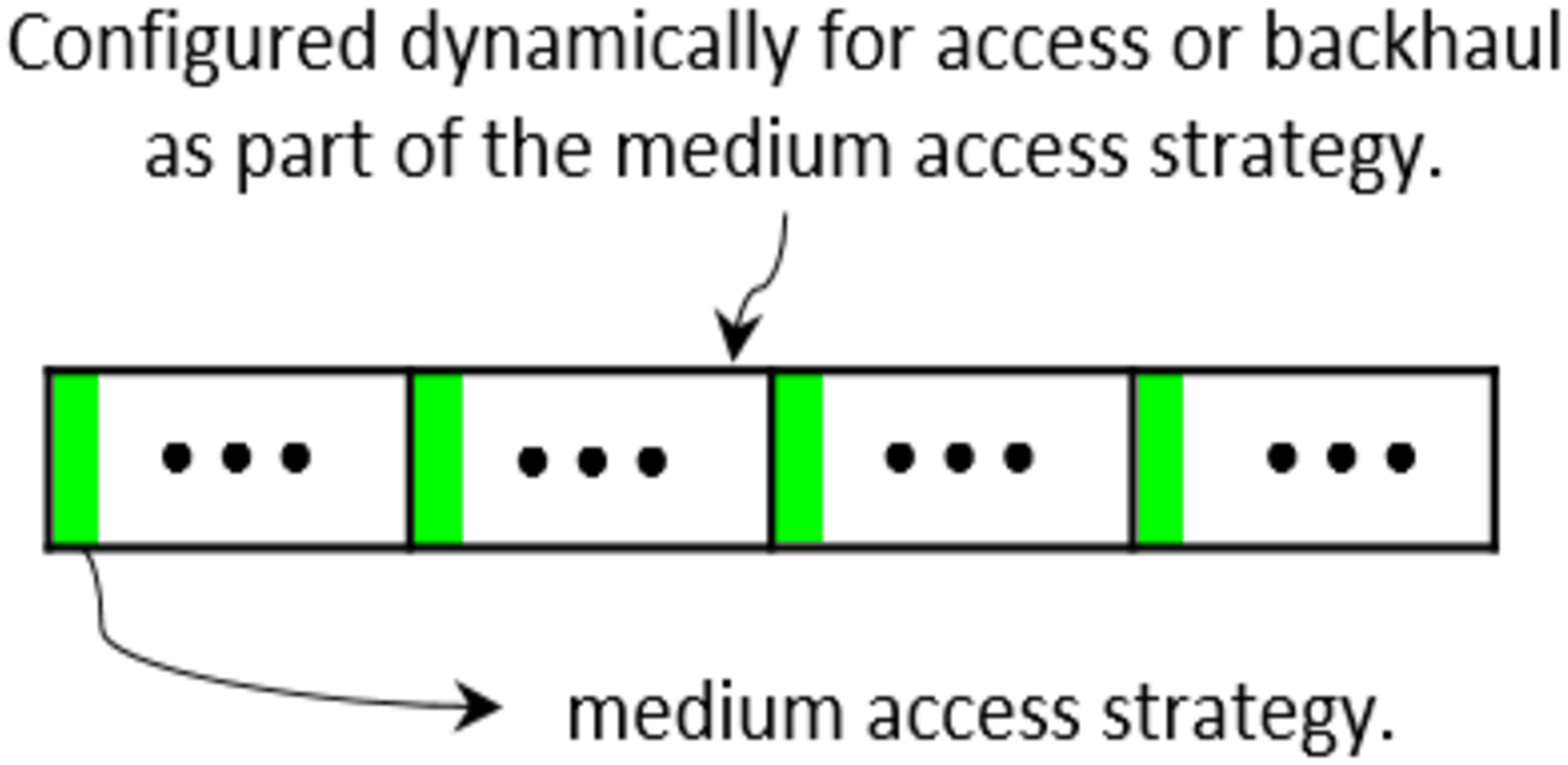}
        \caption{Proposed Strategy}
    \end{subfigure}
    \caption{An illustration of three considered resource allocation strategies for IAB node.}
    \label{fig:fig3}
\end{figure*}

We proposed two LBT-based medium access strategies and aimed to compare them with a modified LTE time-division duplexing (TDD)-based strategy. Investigated three strategies are illustrated in Fig.~\ref{fig:fig3}, and the same has been used for system evaluation in a later section. We briefly discussed the three strategies as below.

\subsubsection{Baseline Strategy}
In this strategy, we adopted the modified LTE TDD frame structures as the baseline, which notably allows the dynamic adaptation of the TDD frame pattern based on available traffic at the node. The frame consists of 10 sub-frames, which are pre-configured for access ($A$) or backhaul ($B$) resources. This strategy offers seven different patterns to statically configure the sub-frames as $A$ or $B$, as shown in Fig.~\ref{fig:fig3}a. The 'special' sub-frame '$S$' can dynamically be configured to either access or backhaul transmission.

\subsubsection{Probabilistic Strategy}
In this strategy, illustrated in Fig.~\ref{fig:fig3}b, we assume that an arriving flow of session requests is divided in two. The probability that a session is routed to the first resource pool is $\delta$, i.e., for backhaul link. With the complementary probability, $1- \delta$, the session requests resources from the second pool, i.e., for access link. A session is dropped if no sufficient resources are available in the selected resource pool. Thus, the system throughput can be calculated using (\ref{eq:iab_access_prob}) is determined as 
\begin{equation}
    \mathcal{R}_{IAB} = \mathbb{P}_\Delta \bigg(\delta \frac{\mathbb{E}[P_k] P_s}{\mathbb{E}[T]} t + (1 - \delta) \frac{\mathbb{E}[P_k] P_s}{\mathbb{E}[T']} t'\bigg) 
\end{equation}
where $\mathbb{E}[P_k]$ is average packet size, $P_s$ denotes the successful transmission rate in a random slot time, $t \in \{0, 1\}$ denoted as bachaul link, $t'$ is complement of $t$ denoted as access link, $\mathbb{E}[T]$ and $\mathbb{E}[T']$ is the average length of a time slot for backhaul and access link, respectively.

\subsubsection{Proposed Strategy}
In this strategy, the scheduling can be achieved using a weighted fair variable, by setting the scheduling weights for access or backhaul link to $\zeta_a$ or $\zeta_b = \frac{1}{\mu}$, where $\mu$ is the reported SINR of access or backhaul link. We assume the SINR can be reported on either PUCCH or PUSCH or on both physical channels for simplicity. In this strategy, in contrast to the other two, the scheduling decision primarily depends on the medium status (idle or busy). If the medium is idle, then only the particular slot is allocated either to access or backhaul, depending upon the SINR status of both of the links. Otherwise, it does not allocate resources at all. As shown in Fig.~\ref{fig:fig3}c, we first assess the status of the channel then schedule the access or backhaul link for a period of maximum occupancy time (COT), as per the regulatory requirements.

\section{Performance Evaluation and Analysis}\label{sec:per_eval}
In this section, we first explained the system model used for the performance evaluation and then demonstrate the performance of the discussed strategies through simulations.

\subsection{Experiment Setup}
\begin{figure}[t]
\centerline{\includegraphics[width=\columnwidth]{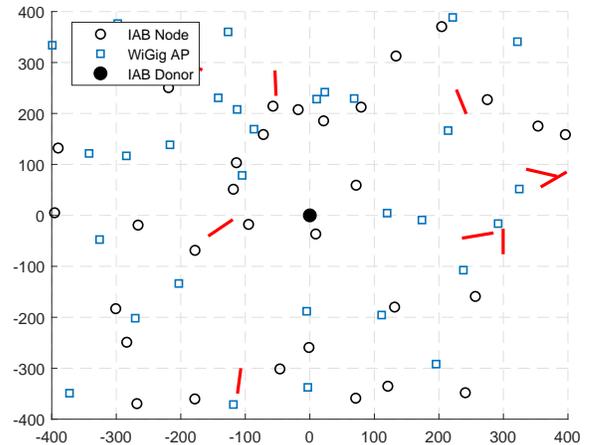}}
\caption{Simulated IAB environment.}
\label{fig:simnet}
\end{figure}

To evaluate the performance, we conducted a simulation study using MATLAB and compared all three schemes. The IAB node and WiGig APs are deployed with three sectors; each sector has a varying number of UEs. UEs are randomly deployed within the transmitting area of each sector. The simulation parameters comply with the latest 3GPP simulation methodology guideline as in \cite{3GPP_ch_mod}. We simulate the throughput and show the cumulative distribution function (CDF) of the average perceived cell throughput and user throughput for all the users in the system. Unless state otherwise, the parameters we used during the simulations are listed in TABLE~\ref{tbl:sys_par}. All the statistical results are averaged over several independent runs. Fig.~\ref{fig:simnet} shows the simulated network scenario. The number of gNBs (i.e., IAB nodes) and Wi-Fi APs are deployed with a 40:60 ratio with an IAB-donor at the center of the network. To operate the IAB not in a fair and friendly manner to WiGig, by not impacting WiGig’s performance more than another WiGig device would do, we configured both the network parameter with equal values.

\begin{table}[t]
\centering
\caption{Simulation Parameters}
\begin{IEEEeqnarraybox}[\IEEEeqnarraystrutmode\IEEEeqnarraystrutsizeadd{2pt}{1pt}]{v/l/v/c/v/c/v}
\IEEEeqnarrayrulerow\\&\mbox{\bf Parameter}&&\mbox{\bf IAB Node}&&\mbox{\bf Wi-Fi}&\\\IEEEeqnarraydblrulerow\\
\IEEEeqnarrayseprow[3pt]\\&\mbox{Carrier frequency}&&$28 GHz$\hfill&&$28 GHz$\hfill&\IEEEeqnarraystrutsize{0pt}{0pt}\\
\IEEEeqnarrayseprow[3pt]\\\IEEEeqnarrayrulerow\\\IEEEeqnarrayseprow[3pt]\\&\mbox{System bandwidth}&&$100 MHz$\hfill&&$100 MHz$\hfill&\IEEEeqnarraystrutsize{0pt}{0pt}\\
\IEEEeqnarrayseprow[3pt]\\\IEEEeqnarrayrulerow\\\IEEEeqnarrayseprow[3pt]\\&\mbox{Network layout}&&$Outdoor$\hfill&&$Outdoor$\hfill&\IEEEeqnarraystrutsize{0pt}{0pt}\\
\IEEEeqnarrayseprow[3pt]\\\IEEEeqnarrayrulerow\\\IEEEeqnarrayseprow[3pt]\\&\mbox{User density}&&$20 users/cell$\hfill&&$20 users/cell$\hfill&\IEEEeqnarraystrutsize{0pt}{0pt}\\

\IEEEeqnarrayseprow[3pt]\\\IEEEeqnarrayrulerow\\\IEEEeqnarrayseprow[3pt]\\&\mbox{CCA threshold}&&$-63 dBm$\hfill&&$-63 dBm$\hfill&\IEEEeqnarraystrutsize{0pt}{0pt}\\
\IEEEeqnarrayseprow[3pt]\\\IEEEeqnarrayrulerow\\\IEEEeqnarrayseprow[3pt]\\&\mbox{Transmit power}&&$23 dBm$\hfill&&$23 dBm$\hfill&\IEEEeqnarraystrutsize{0pt}{0pt}\\
\IEEEeqnarrayseprow[3pt]\\\IEEEeqnarrayrulerow\\\IEEEeqnarrayseprow[3pt]\\&\mbox{Antenna gain}&&$10 dBi$\hfill&&$10 dBi$\hfill&\IEEEeqnarraystrutsize{0pt}{0pt}\\
\IEEEeqnarrayseprow[3pt]\\\IEEEeqnarrayrulerow\\\IEEEeqnarrayseprow[3pt]\\&\mbox{Pathloss parameter ($PL$),}&&\mbox{$\alpha = 72.0$}\hfill&&\mbox{$\alpha = 72.0$}\hfill&\IEEEeqnarraystrutsize{0pt}{0pt}\\
\IEEEeqnarrayseprow[3pt]\\&\mbox{$PL = \alpha + 10 \beta \log_{10} (d)$ [dB]}&&\mbox{$\beta = 2.92$}\hfill&&\mbox{$\beta = 2.92$}\hfill&\IEEEeqnarraystrutsize{0pt}{0pt}\\
\IEEEeqnarrayseprow[3pt]\\\IEEEeqnarrayrulerow\\\IEEEeqnarrayseprow[3pt]\\&\mbox{Tx and Rx beamwidth}&&\mbox{$45^\circ~|~45^\circ$}\hfill&&\mbox{$45^\circ~|~45^\circ$}\hfill&\IEEEeqnarraystrutsize{0pt}{0pt}\\
\IEEEeqnarrayseprow[3pt]\\\IEEEeqnarrayrulerow
\end{IEEEeqnarraybox}
\label{tbl:sys_par}
\end{table}

\subsection{Simulation Results}
\begin{figure}[t]
    \centering
    \includegraphics[width=\columnwidth]{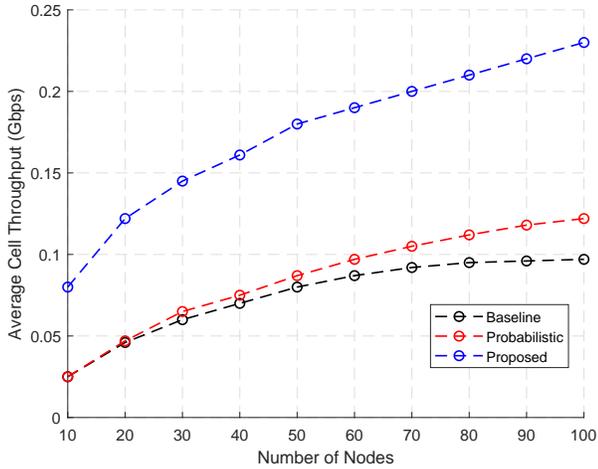}
    \caption{Average cell throughput.}
    \label{fig:cell_tput}
\end{figure}

Fig.~\ref{fig:cell_tput} shows the average cell throughput during channel access. In all three strategies, the cell throughput appears to increase with an increase in the number of nodes monotonically. However, we see a minor throughput gap between the baseline and probabilistic strategy, whereas the Proposed strategy outperformed the other two. Additionally, in the baseline and probabilistic strategy, we observed as the number of APs and nodes increases, the interference increases, which led to a decrease in sumrate. In contrast, the cell throughput appears to increase with nodes due to clear channel assessment monotonically.

\begin{figure}[t]
    \centering
    \includegraphics[width=\columnwidth]{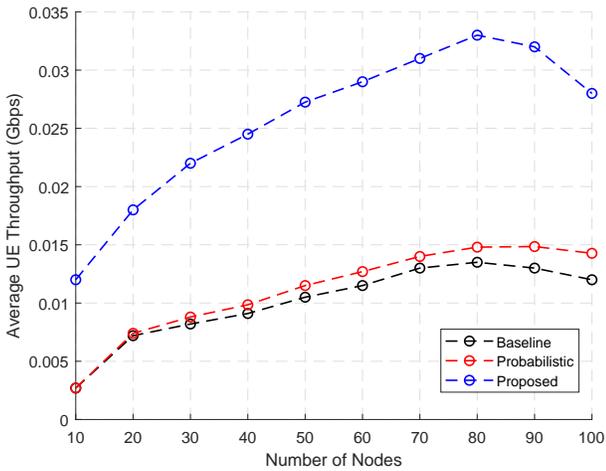}
    \caption{Average UE throughput.}
    \label{fig:ue_tput}
\end{figure}

Fig.~\ref{fig:ue_tput} depicts the average UE throughput results. Like cell throughput, in all three strategies, the UE throughput also appears to monotonically increases with an increase in the number of nodes. However, we see the throughput is exhibited when the UE number of nodes reaches 80. The throughput is affected because of the number of interfering signals due to incorrect LBT interpreting the packet loss at the UE. These unwanted UE-specific interfering signals begin to emerge when the number of nodes reaches 80, degrading the UE throughput.

In the proposed strategy, we first assess the interference level at the receiver and then schedule the transmission. However, in the other two strategies, we first schedule the transmission then access the medium. Thus, as shown in Fig.~\ref{fig:inf}, in these two strategies, incorrect LBT plays a major role, which leads to interference and affects the throughput.

\section{Conclusion}\label{sec:conclusion}
We studied IAB networks from both unlicensed-based spectrum access and performance points of view. Besides, we studied three resource allocation strategies for the IAB scenario in unlicensed/shared spectrum bands with directional transmissions and receptions. The key idea of this work is to perform the feasibility study in the IAB network operating on mmWave unlicensed spectrum. Our study concludes that our proposed medium access-based resource allocation outperforms the LTE TDD-based baseline strategy and our proposed probabilistic-based resource allocation.

\bibliographystyle{IEEEtran}
\bibliography{conference_51064X}

\end{document}